\documentclass[aps,preprint,a4paper,showpacs,showkeys,superscriptaddress]{revtex4-1}
\ifx\pdfoutput\undefined
\usepackage[usenames]{color} 
\usepackage[dvips]{graphicx}
\else
\usepackage{color} 
\definecolor{BrickRed}{rgb}{0.85,0.15,0.25}
\definecolor{MidnightBlue}{rgb}{0,0.45,0.85}
\definecolor{ForestGreen}{rgb}{0,0.85,0.45}
\usepackage{graphicx}
\usepackage{epstopdf}
\usepackage{cancel}
\usepackage{ulem}
\fi
\usepackage{latexsym,amsmath,amssymb}
\allowdisplaybreaks

\usepackage{umoline}
\newsavebox\CBox

\begin{document}


\title{Black Hole Complementarity in Gravity's Rainbow}

\author{Yongwan Gim}%
\email[]{yongwan89@sogang.ac.kr}%
\affiliation{Department of Physics, Sogang University, Seoul 121-742,
  Republic of Korea}%

\author{Wontae Kim}%
\email[]{wtkim@sogang.ac.kr}%
\affiliation{Department of Physics, Sogang University, Seoul 121-742,
  Republic of Korea}%

\date{\today}

\begin{abstract}
To see how the gravity's rainbow works for black hole complementary,
we evaluate the required energy for duplication of information
in the context of black hole complementarity by calculating the critical value of the rainbow parameter in the certain class of the rainbow Schwarzschild black hole.
The resultant energy can be written as the well-defined limit
for the vanishing rainbow parameter which characterizes the deformation of the relativistic dispersion relation
in the freely falling frame.
It shows that the duplication of information in quantum mechanics could
not be allowed
below a certain critical value of the rainbow parameter; however, it might be possible
above the critical value of the rainbow parameter, so that the consistent
formulation in our model
requires additional constraints or any other resolutions for the latter case.
\end{abstract}



\maketitle


\section{Introduction}
\label{sec:intro}
The discovery of Hawking's thermal radiation from a black hole
\cite{Hawking:1974rv,Hawking:1974sw} has raised the information loss paradox
when the black hole evaporates completely \cite{Hawking:1976ra}.
This paradox could be solved for the fixed observer outside the horizon if the Hawking radiation carried
the information of infalling matter state so that the quantum-mechanical evolution
could be unitary.
For the consistency between general relativity and
quantum mechanics,
it requires black hole complementarity
\cite{Susskind:1993if, Susskind:1993mu, Stephens:1993an} which tells us that
the infalling observer (Alice) sees nothing strange when passing through
the horizon based on the equivalence principle while the
external fixed observer (Bob) outside
the horizon sees the information reflected at the stretched horizon based on quantum mechanics.
The two observers of Alice and Bob can only detect the information inside and
outside the black hole, respectively, but
never both simultaneously.
Bob measures the information of the infalling state
during a certain time which amounts to
at least the Page time \cite{Page:1993df,Page:1993wv}
and then jumps into the horizon, he cannot receive the message
from Alice since the available time to send the message to him
is too short. It means that
the required energy derived from the Heisenberg's uncertainty principle exceeds the mass of the black hole
so that black hole complementarity turns out to be valid \cite{Susskind:1993mu}.
Moreover, black hole complementarity is marginally safe for the scrambling time \cite{Hayden:2007cs, Sekino:2008he},
and the recent application of the membrane paradigm appears in Ref. \cite{Fischler:2015cma}.

On the other hand, it has been claimed that there is a puzzle referred
to as the firewall paradox of quantum black holes based on the monogamy principle in quantum mechanics
and semiclassical quantum field theory~\cite{Almheiri:2012rt,Almheiri:2013hfa}.
The consistency of quantum mechanics
requires that the freely falling observer should burn up when crossing the horizon in virtue of the firewall, which
means that the probing inside the black hole is forbidden.
A similar prediction called the energetic curtain around the black hole
from different assumptions was also studied in Ref.~\cite{Braunstein:2009my}.
Subsequently, much attention
has been paid to study the firewall issue along with not only resolutions from various viewpoints
but also including some arguments of the absence of the firewall \cite{Bousso:2012as, Nomura:2012sw,Susskind:2012rm,
Hossenfelder:2012mr,
Page:2012zc,Giddings:2012gc, Mathur:2012jk, Chowdhury:2012vd, Bena:2012zi, Avery:2012tf, Nomura:2012cx,  JACOBSON:2013ewa, Kim:2013fv,Giddings:2013kcj,Maldacena:2013xja,Giddings:2013vda,Almheiri:2013wka,
Hutchinson:2013kka, Page:2013mqa}.
Essentially, most issues on black hole complementarity
are related to the interplay between the Hawking radiation from
the quantum mechanics and the geometry from general relativity.
Now, one might consider additional element that is the
interaction between the infalling apparatus and Hawking radiation plus
geometry.
In fact, the effect of the infalling apparatus
on the gravity background is assumed to be small compared to the black hole mass $M$
for the definite explanation \cite{Susskind:1993mu}.
For a further test for black hole complementarity,
one might want to find a convenient
setting to take into account
this effect of the test particle or the apparatus on the black hole.

On the other hand, it is worth noting that,
based on not only experimental explanation for the threshold anomalies
in ultra high cosmic rays
and Tev photons \cite{AmelinoCamelia:1997gz, AmelinoCamelia:1997jx, Colladay:1998fq, Coleman:1998ti, AmelinoCamelia:1999wk,AmelinoCamelia:2000zs, Jacobson:2001tu, Jacobson:2003bn}
but also theoretic points of view
in the semi-classical limit of loop quantum gravity \cite{Gambini:1998it, Alfaro:2001rb, Sahlmann:2002qk, Smolin:2005cz}, there have been a number of applications
in the framework of gravity's rainbow in order to explore
some effects of a test particle on black holes and cosmology
\cite{Galan:2004st, Hackett:2005mb, Aloisio:2005qt, AmelinoCamelia:2005ik, Ling:2005bp, Galan:2006by, Ling:2006az, Ling:2006ba, Amelino-Camelia:2013wha, Barrow:2013gia, Ling:2008sy,  Garattini:2011hy, Girelli:2006fw, Garattini:2011fs, Garattini:2011kp, Liu:2007fk, Peng:2007nj, Li:2008gs,Ali:2014xqa, Awad:2013nxa, Garattini:2014rwa}.
Some years ago,
Amelino-Camelia proposed that the modified dispersion relation could come from a deformation of the
classical relativity called the doubly special relativity
\cite{AmelinoCamelia:2000ge, AmelinoCamelia:2000mn}
which is the extended version of the Einstein's special theory of relativity
in which the Plank length is also required to be invariant under any inertial frames
along with the invariant speed of light. And then
the most common illustration was presented
by Magueijo and Smolin  \cite{Magueijo:2001cr, Magueijo:2002am,Magueijo:2002xx},
which indeed gives rise to the modified dispersion relation.
This notion was promoted to the curved spacetime in which
the general spacetime background felt by a test particle would depend on
its energy so that the energy of the test particle deforms the background geometry.
Recently, the information loss paradox in gravity's rainbow was discussed in Ref.~\cite{Ali:2014cpa}
by obtaining the finite coordinate time
for the asymptotic observer along the geodesic curve, and
some subtleties of specification of the event horizon were pointed out.

In this work,
we would like to study information loss paradox in the rainbow Schwarzschild black hole
in the context of duplication of information.
For this purpose, we will take into account
the effect of probing particles on the black hole
by employing the special class of the rainbow metric, and calculate
the required energy
in the freely falling frame for Alice to send the message to Bob who
jumped into the black hole at the Page time
whether the energy of quanta will be able to become an enormous scale or not
by calculating the critical value of the rainbow parameter
in the gravity's rainbow.
Let us first recapitulate
the formalism of gravity's rainbow and introduce the rainbow Schwarzschild metric incorporated with
the effect of the test particle in the self-contained manner in section II.
Next, assuming the Heisenberg uncertainty relation, the energy associated with the
proper time will be calculated
on the particular rainbow functions \cite{AmelinoCamelia:2008qg}
in order for the definite illustration in section III.
The deformed metric from the test particle effect will be characterized
by a rainbow parameter $\eta$.
The energy uncertainty will reproduce the well-known limit
presented by Susskind and Thorlacius
for the vanishing rainbow parameter \cite{Susskind:1993mu}.
Furthermore, it turns out that
duplication of information by Alice could be still impossible
below a certain critical rainbow parameter
while the duplication of information might be possible
above the critical rainbow parameter.
The latter case is inconsistent with the assumptions of black hole complementarity
and thus no-cloning theorem of quantum information will be violated.
Finally, conclusion and discussion will be given in section IV.

\section{Rainbow Gravity}
\label{sec:rainbow}
Let us start with
the modified dispersion relation in the
doubly special relativity \cite{AmelinoCamelia:2000ge, AmelinoCamelia:2000mn} which
makes both the speed of light and the Planck length invariant
under the non-linear Lorentz transformation in the momentum space,
which is compactly given as \cite{Magueijo:2001cr, Magueijo:2002am}
\begin{equation}\label{MDR}
f(E/E_p)^2 E^2-g(E/E_p)^2 c^2 p^2 = m^2 c^4,
\end{equation}
where the Planck energy is $E_p=\sqrt{c^5\hbar/G}$, and $E$ and $m$ are the energy and the mass of the test particle,
respectively.
In this formulation, the rainbow functions $f(E/E_p)$ and $g(E/E_p)$
satisfy  $\lim_{E\rightarrow 0} f =1$ and $\lim_{E\rightarrow 0} g =1$
so that the modified dispersion is reduced to the conventional one when the energy
of the test particle vanishes.
The metric based on the modified equivalence principle could also be expressed in terms of
a one-parameter family of orthonormal frame fields \cite{Magueijo:2002xx},
\begin{equation}
g^{\mu\nu}(E/E_p)= \eta^{ab} e_a^\mu (E/E_p) e_b^\nu (E/E_p),
\end{equation}
where the energy dependent frame fields are written as
$e_0(E/E_p)=f^{-1}(E/E_p)\tilde{e}_0$ and $e_i(E/E_p)=g^{-1}(E/E_p)\tilde{e}_i$
and $\tilde{e}$ is the ordinary energy-independent one.
According to this fact, the Einstein field equations are also redefined as
\begin{equation}
\label{meq}
G_{\mu\nu} (E/E_p)=8\pi G(E/E_p) T_{\mu\nu}(E/E_p),
\end{equation}
where $G(E/E_p)$ is the energy dependent Newton constant which
becomes the conventional Newton constant as
$G=G(0)$ for $E=0$.
From Eq. \eqref{meq}, the energy-dependent Schwarzschild solution can be obtained as
 \cite{Magueijo:2002xx}
\begin{equation}\label{metricrain}
ds^2=-\frac{1}{f^2(E/E_p)}\left(1-\frac{2G(0) M}{r}\right)dt^2+\frac{1}{g^2(E/E_p)\left(1-\frac{2G(0) M}{r}\right)}dr^2+\frac{r^2}{g^2(E/E_p)}d\Omega^2,
\end{equation}
where the spacetime coordinates are chosen as the energy-independent coordinates.
In the next section, after deriving the generic form of the energy relation,
we will choose specific rainbow functions \cite{AmelinoCamelia:1997gz}
which can also be obtained from loop quantum gravity approach \cite{Gambini:1998it, Alfaro:2001rb, Sahlmann:2002qk,  Smolin:2005cz} as
\begin{equation}\label{fandg}
f(E/E_p)=1, \qquad g(E/E_p)=\sqrt{1-\eta \left(\frac{E}{E_p}\right)^n},
\end{equation}
where $\eta$ is the rainbow parameter and $n=2$ is chosen in order for analytic calculations.

\section{$\eta$-dependent energy for duplication of information}
Based on the calculations in Refs.~\cite{Susskind:1993if,Susskind:1993mu},
we can calculate the required energy for two copies of information
on the Schwarzschild black hole
which is described by using the Kruskal–Szekeres coordinates of
$ds^2=-32M^3 r^{-1} e^{-r/2M} dU dV$ where $U=- e^{-(t- r^*)/4M},~V=e^{(t+ r^*)/4M}$ and $r^*=r+2M \ln(| r- 2M |/2M)$.
For convenience, the constants are set to $G=\hbar=c=k_B=1$; however,
these will be recovered as necessary.
After Alice passes through the stretched horizon, Bob will essentially begin to
observe the information through the Hawking radiation at the Page time
$t_{\text{Page}} \sim M^3$ \cite{Page:1993df}.
Let us assume that Alice passes through the stretched horizon
at $V_A=1$, then Bob should go through the stretched horizon at least after
the Page time, so $V_B\sim e^{M^2}$.
To send the information by Alice before Bob arrives at the curvature singularity,
Alice should send the information before $U_A=1/V_B \sim e^{-M^2}$,
where the curvature singularity appears at $UV=1$.
Next, the interval of the proper time $\Delta \tau$
which is nothing but the free-fall time for Alice near $V_A=1$ is given as
$\Delta \tau \sim M e^{-M^2}$.
By using the Heisenberg's energy-time uncertainty principle of
$\Delta \tau \Delta E \geq 1/2$,
one can eventually obtain $\Delta E \sim M^{-1} e^{M^2}$
which is indeed larger
than the mass of black hole, and
the information should be encoded into the radiation with super-Planckian scale of the quanta.
Therefore, Bob cannot see the duplication of Alice's information physically,
so that black hole complementarity is valid.

Let us now calculate the energy for the duplication of information
by using the rainbow Schwarzschild metric \eqref{metricrain} along the
argument in Ref. \cite{Susskind:1993mu}.
First, the metric \eqref{metricrain} is written as
\begin{equation}\label{kruskalrain}
 ds^2 = -\frac{4 r_H^3 }{g^2 r}e^{-r/r_H}dUdV,
\end{equation}
in terms of the rainbow Kruskal–Szekeres coordinates defined as
 $U = -e^{-((g/f)t- r^*)/2r_H},~V =  e^{((g/f)t+r^*)/2r_H}$ and $r^*=r+r_H \ln(| r-r_H|/r_H)$.
The rainbow metric and the conformal transformations
are reduced to the conventional ones for $f=g=1$ where the probing energy $E$ goes to zero.

To get the information retention time for the rainbow black hole,
we consider the Stefan-Boltzmann law written as \cite{Page:1993wv}
\begin{equation}\label{SBlaw}
 \frac{dM}{dt}= -A\sigma T^4,
\end{equation}
where $A$ denotes the area of a black body and $\sigma=\pi^2k_B^4/(60 \hbar^3c^2)$ is
the Stefan-Boltzmann constant, and the Hawking temperature $T$
is also calculated as
\begin{equation}\label{HawkingT}
T= \frac{\kappa_{\rm H}}{2\pi} = \frac{1}{8\pi G M}\frac{g(E/E_p)}{f(E/E_p)},
\end{equation}
where $\kappa_H$ is the surface gravity at the horizon.
Using the
explicit form of the rainbow functions \eqref{fandg}, the black hole temperature can be written as
 \cite{Ali:2014xqa}
\begin{equation}
\label{tem}
T_{\rm H}= \frac{1}{8\pi G M}\sqrt{1-\eta\left(\frac{E}{E_p}\right)^n}.
\end{equation}
In order to eliminate the dependence of the particle energy in the black temperature
\eqref{tem},
one can use the heuristic method in Ref. \cite{Adler:2001vs}.
It shows that the Heisenberg uncertainty principle gives a relation between
the momentum of the Hawking particle
$p$ emitted from the black hole and the mass of  black hole $M$
as $p= \Delta p \sim  1/(2 G M) $ by regarding $\Delta x  \sim 2GM$  \cite{Adler:2001vs}.
Next, for a definite illustration, we choose $n=2$.
So the energy for the massless particle can be easily
solved as $E =  E_p/({\sqrt{\eta+4 G^2 E_p^2 M^2}})$.
Plugging this into the temperature \eqref{tem}, one can obtain
the rainbow Hawking temperature as \cite{Gim:2014ira}
\begin{equation}\label{T2}
T_{\rm H}=  \frac{1}{8\pi G M}\sqrt{\frac{4G M^2}{4G M^2+\eta}},
\end{equation}
where it reproduces the well-known Hawking temperature for the large black hole.
Next, from the Stefan-Boltzmann law \eqref{SBlaw},
the Page time when the black hole has emitted half of its initial Bekenstein-Hawking entropy
can be calculated as \cite{Page:1993wv}
\begin{equation}\label{tpage}
t_{\text{Page}} \sim M^3  + \mathcal{O}\left(\eta\right),
\end{equation}
where the black hole was assumed to be very large even after the Page time.

\begin{figure}[pt]
  \begin{center}
  \includegraphics[width=0.55\textwidth]{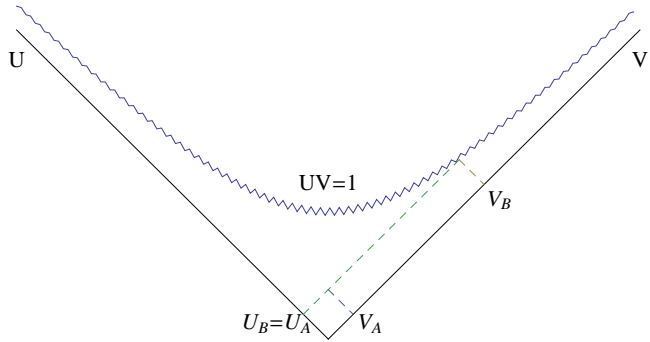}
  \end{center}
  \caption{
  The wiggly curve denotes the black hole singularity where $UV=1$. Alice passes through the horizon at $V_A$
  and then Bob also jumped into the horizon at $V_B$ just after the Page time. Alice should send the signal to Bob before
  the message hits the singularity.}
  \label{fig:penrose}
\end{figure}

Based on gedanken experiments, let us now
suppose that
Alice passes through the stretched horizon
at $V_A(t_A,r^*_A)=1$
in Fig. \ref{fig:penrose}
and then
after a period of the Page time \eqref{tpage} outside the black hole Bob will jump into the stretched horizon
at $V_B(t_B,r^*_B)$
with the information which has been gathered from Hawking radiation,
$i.e.$,  $t_B=t_A+t_{\text{Page}}$.
It means that Bob starts to move toward the horizon
at
$V_B =  e^{((g/f)t_B+r^*_B)/2r_H}
=  e^{((g/f)(t_A+t_{\text{Page}})+r^*_A)/2r_H}
= e^{((g/f)t_{\text{Page}})/(2r_H)}V_A$.
Note that $r_A^*=r_B^*$ and $V_A=1$.
Thus Alice should send her messages before $U_A=1/V_B= e^{- (g/f)M^2}$,
so that from Eq. \eqref{kruskalrain}
 the proper time measured by Alice near the horizon $r  \sim r_H$ can be written as
\begin{align}
\Delta \tau^2 &=\frac{4r_H^2}{g^2}e^{-1}(U_A-0)\Delta V_A \\
  &\sim \frac{1}{g^2} M^{2} e^{-(g/f)M^2}\label{Dtau},
\end{align}
where we assumed that $\Delta V_A$ is a nonvanishing finite value near $V=1$
\cite{Susskind:1993mu}.

Next, one can find the appropriate energy-time uncertainty relation in the local
rainbow inertial frame. For this purpose, we assume that the usual
Heisenberg uncertainty principle is valid
in the local inertial frame
where there is no rainbow effect.
Then the uncertainty relation between the energy and time
in our case should be written as
  \begin{equation}\label{EtMDR}
\Delta \tau \Delta E \geq \frac{1}{2f}
\end{equation}
by taking into account the rainbow effect.
It tells us that the energy uncertainty becomes $\Delta E \sim g/f $ since
the proper time is proportional to $1/g$ from Eq. \eqref{Dtau},
which is compatible with the structure of the modified dispersion relation \eqref{MDR}.
Consequently, using Eqs. \eqref{Dtau} and \eqref{EtMDR},
the energy uncertainty is obtained as
\begin{equation}\label{DE1}
 \Delta E^2 \sim \frac{M_p^4c^4\left(1-\eta \left(\frac{\Delta E}{E_p}\right)^2\right)}{M^2} e^{\frac{M^2}{M_p^2}\sqrt{1-\eta  \left(\frac{\Delta E}{E_p}\right)^2}}
\end{equation}
by recovering dimensional constants.
Note that in the absence of the rainbow effect of $\eta \to 0$,
the energy uncertainty is consistent with
the standard result of the Susskind-Thorlacius limit of $\Delta E \sim (M_p^2c^2/ M) e^{M^2/M_P^2}$
 whose value could be beyond the Planckian scale for the
large black hole \cite{Susskind:1993mu}.
\begin{figure}[pt]
  \begin{center}
  \includegraphics[width=0.55\textwidth]{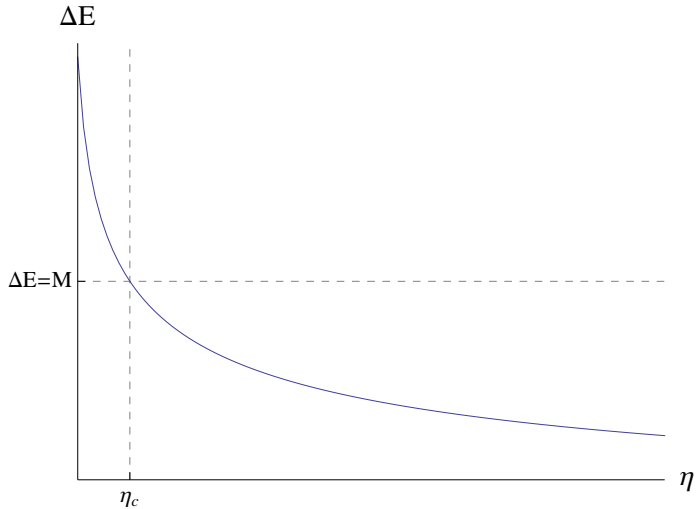}
  \end{center}
  \caption{
    The energy uncertainty vs the rainbow parameter is plotted by setting M=2 for convenience.
    It is indeed monotonically decreasing function for the arbitrary M.}
  \label{fig:DEeta}
\end{figure}
From Eq. \eqref{DE1},
one can see that $d(\Delta E(\eta))/d\eta$ for an arbitrary $M$ is always negative,
so that it is monotonically decreasing as shown in Fig. \ref{fig:DEeta}.
Instead of solving the closed form of Eq. \eqref{DE1} perturbatively,
we provide a criterion for information cloning, and define a
critical case
where the energy uncertainty at the Page time amounts to the black hole mass of $\Delta E \sim M$,
which is given as
\begin{equation}\label{eta}
 \eta_c= \frac{M^4-4W^2(Z)}{M^6}
\end{equation}
by means of the Lambert W function defined as $Z=W(Z)e^{W(Z)}$ where $Z=M^4/2$.
Therefore, no-cloning theorem of quantum information is valid as long as $\eta \ll \eta_c$
since the required energy of the quanta for the cloning exceeds the mass of the black hole
like the result in Ref. \cite{Susskind:1993mu},
while it might be violated for  $\eta \gg \eta_c$.

\label{sec:Dis}
\section{Conclusion and Discussion}
The required energy in
the freely falling frame for Alice to send the message to Bob who
jumped into the black hole at the Page time was calculated
in the rainbow Schwarzschild black hole
in order to investigate how black hole complementarity
works in the rainbow gravity. It tells us that
the rainbow parameter should be much less than the
critical rainbow parameter to maintain the unitarity
in quantum mechanics so that black hole complementarity can be safe.
If the rainbow parameter were much larger than the critical one, then
the energy uncertainty would be made smaller than the Planckian scale of the
energy. However, it does not mean that the generic rainbow gravity can save black hole
complementarity since the present calculation is based on the particular rainbow
functions.

As for the case to violate the unitarity, it is necessary to introduce
additional devices or resolutions to protect unitarity in the rainbow Schwarzschild black hole.
So we would like to comment on some
various speculations to resolve this problem.
i) A certain
 rainbow-improved black hole complementarity might exist and it
 should be reduced to the conventional black hole complementarity when the rainbow parameter vanishes.
 ii) The firewall like objects or something else,
for example, the brick wall \cite{Israel:2014eya}
might exist to protect the unitarity for the generic non-vanishing rainbow parameter.
However, it was shown that the brick wall can be eliminated due to the modification of the density of states
by choosing appropriate rainbow functions \cite{Garattini:2009nq}.
iii) As was claimed by Hawking \cite{Hawking:2014tga},
if there were no event horizon behind which information is lost, then
information could be preserved
during the evaporation. It was also stressed that
the event horizons are inappropriate to describe the physical black holes \cite{Visser:2014zqa}.
Recent numerical calculations by
taking into account the quantum back reaction of the geometry
show that a star stops collapsing a finite radius larger than
its horizon \cite{Mersini-Houghton:2014zka, Mersini-Houghton:2014cta}. If this were true,
then it could be analogously applied to this rainbow gravity.
However, it was claimed that
the radial distance between
the event horizon and the apparent horizon is much smaller than
the Planck length for the large black hole \cite{Frolov:2014wja}.
So the information loss paradox seems to be still ongoing issue in the rainbow gravity.

\acknowledgments
We would like to thank M. Eune for exciting discussions and
the anonymous referee for many valuable comments to improve our manuscript.
This work was supported by the National Research Foundation of Korea(NRF) grant funded by the Korea government(MSIP) (2014R1A2A1A11049571).


\end{document}